\documentclass[12pt]{article}

\parskip=\medskipamount            
\def\7#1#2{\mathop{\null#2}\limits^{#1}}        

\newdimen\footheight
\begin{document}
\bibliographystyle{unsrt}

\begin{center}
\textbf{Discovery of the
Color Degree of Freedom\\ in Particle Physics: a Personal Perspective}\\
[5mm]
O.W. Greenberg\footnote{email address, owgreen@umd.edu.}\\
{\it Center for Fundamental Physics\\
Department of Physics\\
University of Maryland\\
College Park, MD~~20742-4111}\\
University of Maryland pp-08-05\\
~\\
\end{center}

\begin{abstract}
I review the main features of the color charge degree of freedom 
in particle physics, sketch the paradox in the early quark model that led to color, 
give a personal perspective on
the discovery of color and describe the introduction of the gauge theory of color.
\end{abstract}

\section{Introduction}
Our present conception of the nature of elementary particles
includes fractionally charged quarks that
carry a hidden 3-valued charge degree of freedom, ``color,'' as
fundamental constituents of strongly interacting particles (hadrons).  
The main features of color are (1) it is a hidden 3-valued charge degree
of freedom carried by quarks, (2) it can be incorporated into an $SU(3)_{color}$
gauge theory, and (3) the hidden color
gauge group commutes with electromagnetism. This third feature requires 
that the electric charges
of quarks are independent of color, which in turn requires the quarks
to have fractional electric charges.

Quarks with fractional electric charges were
introduced by Murray Gell-Mann~\cite{mgm} and, independently, by
George Zweig~\cite{zwe} in 1964. Also in 1964 I introduced color, using
parafermi statistics of order 3~\cite{gre64}. This 3 is the same 3 
as the 3 of $SU(3)_{color}$. 

My work was stimulated 
by the $SU(6)$ theory of Feza G\"ursey and Luigi A.
Radicati~\cite{gur}
in the same year.
G\"ursey and Radicati placed the baryons in the symmetric 3-particle 
representation
of $SU(6)$. This produced a paradox: The spin 1/2 quarks must be fermions,
according to the spin-statistics theorem, and can only occur in 
antisymmetric representations. I
resolved this paradox in 1964~\cite{gre64} by suggesting that quarks 
obey parafermi
statistics~\cite{hsg} of order 3, which allows up to 3 particles to be in a 
symmetric state. As mentioned above, the
3 of the parafermi statistics is the same 3 as the 3 of color $SU(3)$.

Because particles with fractional electric charge had not been observed,
several of the early authors chose models with integer quark charges. Such models
are unacceptable both theoretically and experimentally. In models with
integer quark charges electromagnetism does not commute with color so
that color symmetry is broken. Such a model violates the exact
conservation of color which is a crucial part of QCD.
Integer charges also conflict with experimental
evidence coming from the ratio
$\sigma(e^+ e^- \rightarrow \mathrm{hadrons})
/\sigma(e^+ e^- \rightarrow \mu^+ \mu^-)$ as well as
from the analysis of
jets in high energy hadron collisions. 

I emphasize that there are two independent discoveries connected with the
strong interactions: (1) color as a charge--analogous to electric charge in
electromagnetism, and (2) color as a gauge symmetry--analogous to the $U(1)$
symmetry of electromagnetism.

The gauge symmetry of a theory is intimately connected with the
quantities that are observable in the theory. In the context of
parastatistics if only currents such as
\begin{equation}
[\bar{\psi}(x),\gamma^{\mu} \psi(x)]        \label{b}
\end{equation}
are observable, then the gauge symmetry is $SU(3)$. With additional
observables such as
\begin{equation}
[\psi(x),\gamma^{\mu} \psi(x)]               \label{n}
\end{equation}
the symmetry is $SO(3)$. 

For the parafermi theory of quarks we choose
only baryon number zero currents, so that only the currents of Eq.(\ref{b})
are allowed. Currents such as Eq.(\ref{n}) have non-zero baryon number and
are not allowed. Thus the parafermi theory must be associated with
the symmetry $SU(3)$. We can make this explicit, following Oscar Klein~\cite{kle}, 
by transforming
the Green components of the parafields to sets of normal fields. The
choice of currents with zero baryon number leads to explicit $SU(3)$ symmetry
for the normal quark fields. To summarize, the choices of observables and of
gauge symmetry are directly related.

The parastatistics of H.S. Green cannot be gauged because 
the commutation rules for the Green components with equal values of
the Green index are not the same as the commutation rules for Green
components with unequal values of the Green index.
Kenneth Macrae and I~\cite{gremac}
showed how to modify Green's parastatistics so that it can be gauged by
reformulating parastatistics with Grassmann numbers.
Further, we showed that using
Grassmann numbers that obey an $SU(N)$ ($SO(N)$) algebra leads to an
$SU(N)$ ($SO(N)$) gauge theory.

In summary, the full understanding of color emerged from the work of Greenberg in
1964, and the work of Nambu in 1965 and of Han and Nambu in 1965. As is often the
case in the development of new theories, neither got everything in final
form at the beginning.

\section{Influences leading to the discovery of the hidden 3-valued color 
charge degree of freedom}
Here I describe the disparate influences that led me to introduce
the color charge degree of freedom. In the 1950's and early 1960's
I was struck by the success of very simple ideas in bringing order to the
newly discovered ``strange'' particles. In the same period, under the
influence of my thesis advisor, Arthur Wightman, I was learning
sophisticated mathematical techniques based on quantum field theory. My
PhD thesis on the asymptotic condition in quantum field theory gave a
formalization of the Lehmann, Symanzik, Zimmermann (LSZ) scattering theory.
I used
the theory of operator-valued distributions and gave proofs of properties
of LSZ scattering theory that were mathematically rigorous according to the
standards of that period.

I became interested in the theory of identical
particles as a graduate student in the 1950's.
I wondered why only bosons and fermions occur in 
Nature, as well
as what other possibilities might exist. Although I did not see his paper
at the time, H.S. Green had introduced a generalization of each type of
statistics in 1953. He generalized bose (fermi) statistics to parabose
(parafermi) statistics of integer order $p$. Green replaced the usual bilinear
commutation or anticommutation relations for bose and fermi statistics by
trilinear relations. He found solutions of his trilinear relations using an ansatz. 
Expand a field $q$ in $p$ ``Green'' components that (for the parafermi case)
anticommute when the Green indices are the same but commute when the Green
indices are different,
\begin{equation}
[q^{(\alpha)}(x),q^{(\alpha)~\dagger}(y)]_+=\delta(\mathbf{x}-\mathbf{y}), x^0=y^0,
\end{equation}
\begin{equation}
[q^{(\alpha)}(x),q^{(\beta)~\dagger}(y)]_-=0, x^0 \neq y^0.
\end{equation}
(For the parabose case interchange commutator and anticommutator.) For parafermi (parabose)
statistics of order $p$ at most $p$ identical particles can be in a symmetric (antisymmetric)
state.

As mentioned above, 
Klein gave a recipe for converting fields that anticommute (commute)
into fields that
commute (anticommute). Schematically, his transformation is
\begin{equation}
Q^{(\alpha)}(x)=K_{(\alpha)}q^{(\alpha)}(x),
\end{equation}
\begin{equation}
K_{(\alpha)}|0\rangle=|0\rangle.
\end{equation}
The Klein transformation converts the anomalous anticommutation (commutation) relations
to the normal ones.

Albert M.L. Messiah and I worked together on generalizations
of the usual bose and fermi quantum statistics in 1962-1964 .
We showed that any representation
of the symmetric group for identical particles is compatible with
quantum mechanics in the context of first-quantized quantum theory~\cite{mesgre}. 
We also 
worked out the branching rules for changes in the number of 
identical particles.
We formulated parastatistics without using Green's ansatz (for the case with the
usual Fock vacuum) in the context of second-quantized quantum 
field theory~\cite{gremes}. 
In addition, we derived the selection rules for interactions that change the
number of identical particles. 
This work prepared me to address the paradox in the quark
model of baryons that arose in 1964.

The year 1964 was the crucial year for the discovery of both quarks and color. 
Quarks were suggested independently by Gell-Mann and by Zweig.
Gell-Mann's quarks resembled what we now call current quarks.
Zweig's quarks, which he
called aces, deuces and treys, resembled what we now call constituent quarks. 
In early
1964 when I first heard about rumors of the idea of quarks I wondered why only the
combinations $qqq$ and $\bar{q}q$ occurred in nature. In the original models 
there was no reason for this.

The paradox concerning the quarks in baryons arose in the $SU(6)$ theory of 
G\"ursey and
Radicati. They generalized an idea of Wigner from 1937. 
Wigner had combined the $SU(2)_I$ of isospin
with the $SU(2)_S$ of spin to make an $SU(4)$ and used this $SU(4)$ 
to classify nuclear states
and derive relations for their energy levels. With a larger symmetry group he found
more relations among the energy levels. G\"ursey and Radicati combined the $SU(3)_f$
of the three quark flavors in the original quark model with $SU(2)_S$ to get an
$SU(6)$ that they used to classify particle states.

The $SU(6)$ theory considers a quark as a
\begin{equation}
\mathbf{3} \sim (u,~d,~s) ~\mathrm{in}~ SU(3)_f
\end{equation}
and the spin
\begin{equation}
\mathbf{1/2} \sim ( \uparrow,~ \downarrow) \hspace{.2cm} \mathrm{in} \hspace{.2cm} SU(2)_S.
\end{equation}
G\"ursey and Radicati combined these as a
\begin{equation}
\mathbf{6} \sim (\mathbf{3}, \mathbf{1/2}) \sim
(u_{\uparrow},u_{\downarrow},d_{\uparrow}, 
d_{\downarrow}, s_{\uparrow}, s_{\downarrow}) ~\mathrm{in}~ SU(6).
\end{equation}
We can also decompose the quark under
\begin{equation}
SU(6) \rightarrow SU(3)_f \times SU(2)_S,
\end{equation}
\begin{equation}
\mathbf{6} \rightarrow (u,d,s) \times (\uparrow, \downarrow).
\end{equation}
For the $q \bar{q}$ mesons this works well; we have
\begin{equation}
\mathbf{6} \otimes \mathbf{6}^{\star} = \mathbf{1} + \mathbf{35},
\end{equation}
\begin{equation}
\mathbf{35} \rightarrow (\mathbf{8}, \mathbf{0}) 
+ (\mathbf{1} + \mathbf{8}, \mathbf{1}).
\end{equation}
Here the $\mathbf{8}$ and the $\mathbf{1}$ before the commas are the 
$SU(3)_f$ multiplicities
and the $\mathbf{0}$ and the $\mathbf{1}$ after the commas are the spins of
the particles. The octet of
psuedoscalar mesons,
\begin{equation}
(K^+,K^0,\pi^+,\pi^0,\pi^-,\eta^0, \bar{K}^0,\bar{K}^-),
\end{equation}
was known,
as were the singlet plus octet (or nonet) of vector mesons,
\begin{equation}
(K^{\star +},K^{\star 0},\phi^0,\rho^+\rho^0,\rho^-,\omega^0,
\bar{K}^{\star 0},\bar{K}^{\star -}).
\end{equation}
Both the octet and the nonet fit well in the $SU(6)$ scheme.

The analogous calculation for the $qqq$ baryons requires decomposing the 
product of three $\mathbf{6}$'s
into irreducibles of $SU(6)$,
\begin{equation}
\mathbf{6} \otimes \mathbf{6} \otimes \mathbf{6} 
= \mathbf{56} + \mathbf{70} + \mathbf{70} + \mathbf{20}.
\end{equation}
This $\mathbf{56}$ is the representation that fits the data on the lowlying baryons,
\begin{equation}
\mathbf{56} \rightarrow (\mathbf{8}, \mathbf{1}/\mathbf{2}) 
+ (\mathbf{10}, \mathbf{3}/\mathbf{2}),
\end{equation}
since there is an octet of spin-1/2 baryons ,
\begin{equation}
(p^+, n^0, \Lambda^0, \Sigma^+, \Sigma^0, \Sigma^-, \Xi^0, \Xi^-),
\end{equation}
and a decuplet of spin-3/2 baryons,
\begin{equation}
(\Delta^{++},\Delta^+,\Delta^0,\Delta^-,
Y_1^{\star +},Y_1^{\star 0},Y_1^{\star -},\Xi^{\star 0},
\Xi^{\star -}, \Omega^-).
\end{equation}
G\"ursey and Radicati found a mass formula for these baryons that generalizes the
Gell-Mann--Okubo mass formula for each $SU(3)$ multiplet and also gives a 
new relation between
masses in the octet and the decuplet. 

The $\mathbf{56}$ seemed like a compelling
choice for the baryons in the quark model. However, this leads to a paradox: 
The permutation
properties of the $\mathbf{56}$, $\mathbf{70}$ and $\mathbf{20}$ are respectively 
symmetric,
mixed and antisymmetric. Since the quarks should have spin 1/2,
the spin-statistics theorem~\cite{pau} requires that they should be fermions
and occur in the antisymmetric $\mathbf{20}$ representation. The experimental data
which places the baryons in the symmetric $\mathbf{56}$ representation 
conflicts with
the spin-statistics theorem.

When I came to Princeton in the fall of 1964 there was a lot of excitement about the
G\"ursey-Radicati $SU(6)$ theory. Benjamin W. Lee gave me a preprint of a
paper~\cite{blp} on the
ratio of the magnetic moments of the proton and neutron that he had written with
Mirza A. Baqi B\'eg and Abraham Pais. They had calculated this magnetic moment ratio
using the group theory of $SU(6)$. I translated their result into the
concrete quark model, assuming the quarks obey bose statistics in the visible
degrees of freedom.
Both the result, that the ratio is $-3/2$, and the simplicity of the calculation
were striking. 

Here is my version of that calculation: 
Represent the proton and neutron with spin up as
\begin{equation}
|p_{\uparrow}^+\rangle = \frac{1}{\sqrt{3}}u_{\uparrow}^{\dagger}
(u_{\uparrow}^{\dagger}d_{\downarrow}^{\dagger}
-u_{\downarrow}^{\dagger}d_{\uparrow}^{\dagger})|0\rangle,
\end{equation}
\begin{equation}
|n_{\uparrow}^0\rangle = \frac{1}{\sqrt{3}}d_{\uparrow}^{\dagger}
(u_{\uparrow}^{\dagger}d_{\downarrow}^{\dagger}
-u_{\downarrow}^{\dagger}d_{\uparrow}^{\dagger})|0\rangle.
\end{equation}
The $(u_{\uparrow}^{\dagger}d_{\downarrow}^{\dagger}
-u_{\downarrow}^{\dagger}d_{\uparrow}^{\dagger})$
combination in parentheses serves as a ``core'' that carries
zero spin and isospin, so that the third quark to the left of the
parentheses carries the spin and isospin of the proton or neutron.
The magnetic moment is then the matrix element
$\mu_B = \langle B_{\uparrow}|\mu_3|B_{\uparrow}\rangle$, where
$\mu_3=2 \mu_0 \Sigma_q Q_q S_q, ~Q_q=(2/3, -1/3,-1/3)$, the $2$ is the
g-factor of the quark, $\mu_0$ is the Bohr magneton of the quark and
$Q_q$ are the quark charges in units of the proton charge. With this
setup the magnetic moments can be calculated on one line,
\begin{equation}
\mu_p=2 \mu_0 \frac{1}{3} \{2[\frac{2}{3}\cdot 1+(-\frac{1}{3}) \cdot (-\frac{1}{2})]
+[(-\frac{1}{3}) \cdot (\frac{1}{2})]\}=\mu_0.
\end{equation}
The analogous calculation for the neutron gives
\begin{equation}
\mu_n=-\frac{2}{3}\mu_0.
\end{equation}
The ratio is $\mu_p/\mu_n=-3/2$, which agrees with experiment to $3\%$. 
This leads to an
estimate for the effective mass of the quark in the nucleon, 
$m_N/2.79 \approx 340 MeV/c^2$,
which is consistent with present extimates of the constituent 
masses of the up and down quarks.

Previous calculations of the magnetic moments using pion clouds had failed.
Nobody had realized that the ratio was so simple. In retrospect the calculation
worked better than we would now expect, since it did not take account of 
quark-antiquark pairs and gluons. Nonetheless, for me the success of this simple
calculation was a very convincing additional argument that quarks 
have concrete reality.

The paradox about the placement of the baryons in the $\textbf{56}$
representation of $SU(6)$ was based on the spin-statistics theorem which
states: Particles that have integer spin must obey Bose statistics, and
particles that have odd-half-integer spin must obey Fermi statistics. I knew there
is a generalization of the spin-statistics theorem that was not part of
general knowledge in 1964: Particles that have integer spin must obey
parabose statistics, and particles that have odd-half-integer spin must
obey parafermi statistics~\cite{dgs}. Each family is labeled by an integer $p$;
$p=1$ is the ordinary Bose or Fermi statistics. 

I immediately realized
that parafermi statistics of order 3 would allow up to 3 quarks in the
same space-spin-flavor state without violating the Pauli principle, which
would resolve the statistics paradox. To test this idea I suggested a model in
which quarks carry order-3 parafermi statistics in~\cite{gre64}.
\textbf{This was the
introduction of the hidden charge degree of freedom now called color.}

With this resolution of the statistics paradox I was exhilarated. I felt
that the new charge degree of freedom implicit in the parafermi model would
have lasting value. I became convinced that the quark model and color were
important for the theory of elementary particles. Not everybody shared my enthusiasm.

It is difficult now to grasp the level of rejection of these ideas
in 1964 and even for the next several years.
Quarks were received with skepticism in 1964. Color as a hidden charge carried
by quarks was received with disbelief. 

The reactions of two distinguished
physicists illustrate this skepticism and disbelief. I gave a copy of my
paper to J. Robert Oppenheimer and asked his opinion of my work when I
saw him at a conference about a week later. He said, ``Greenberg, it's
beautiful!,'' which sent me into an excited state. His next
comment, however, ``But I don't believe a word of it.'' brought me down to earth.
In retrospect I have two comments about these remarks of Oppenheimer.
I was not discouraged, because I was convinced that my solution to the statistics
paradox would have lasting value. Nonetheless, I was too intimidated by
Oppenheimer to ask why he did not believe my paper.

Steven Weinberg, who contributed as much as anybody to the standard model
of elementary particles, wrote in a talk on \textit{The Making of the Standard
Model}~\cite{wei} ``At that time [referring to 1967] 
I did not have any faith in the existence of quarks.''

The skepticism about quarks and color can be understood: Quarks were
new. Nobody had ever observed a particle with fractional electric charge.
Gell-Mann himself was ambiguous about their reality. In his paper he wrote
``...It is fun to speculate...if they were physical particles of finite mass
(instead of purely mathematical entities as they would be in the limit of
infinite mass...A search... would help to reassure us of the non-existence
of real quarks~\cite{mgm}.'' To add a hidden charge degree of freedom to the
unobserved fractionally charged quarks seemed to stretch credibility to
the breaking point at that time. 
In addition, parastatistics, with which the new degree
of freedom was introduced, was unfamiliar.

Resolving the statistics paradox was not a sufficient test of color. I needed
new predictions. I turned to baryon spectroscopy to
construct a model of the baryons in which the hidden parafermi (color)
degree of freedom takes care of the required antisymmetry of the Pauli
principle. Then I could treat the quarks as bosons in the visible space,
spin and flavor degrees of freedom, with the parastatistics taking care
of the necessary antisymmetry. I made a table of the excited baryons in
the model using $s$ and $p$ state quarks in the \textbf{56}, $L=0^+$ and
\textbf{70}, $L=1^-$ supermultiplets.

I followed up this work with Marvin Resnikoff in 1967~\cite{greres}.
This work has been
continued by Richard H. Dalitz and collaborators, by Nathan Isgur and Gabriel
Karl and by Dan-Olof Riska and collaborators, among others. The original
fits to the baryons made in 1967 are surprisingly close to the current fits
of 2008.

The only evidence for color from 1964 to 1969 was the baryon
spectroscopy that I proposed in 1964. It was only in 1968 that the first 
rapporteur at an international conference accepted the 
parastatistics model for baryons as the correct model. 
By then the data on baryon spectroscopy
clearly favored the new degree of freedom. In 1969, Steven Adler, John Bell
and Roman Jackiw explained the $\pi \rightarrow \gamma \gamma$ decay rate
using the axial anomaly with colored quarks. This gave the first additional
evidence for quarks.

\section{Introduction of the gauge theory of color}
Explicit color $SU(3)$ was introduced in 1965 by 
Yoichiro Nambu~\cite{nam} and by 
Moo-Young
Han and Nambu~\cite{hannam}. The papers of Nambu and 
of Han and Nambu used 3 dissimilar
triplets
in order to have integer charges for the quarks. This is not correct, both
experimentally theoretically for reasons given above. However this paper
paper includes the statement \textit{``Introduce now eight gauge vector fields
which behave as (1,8), namely as an octet in 
$SU(3)^{\prime \prime}$"}~\cite{hannam}. 
\textbf{This was the
introduction of the gauge theory of color.} 
The $\mathbf{1}$ in the $\mathbf{(1,8)}$ 
refers to
what we now call flavor and is the statement that the gauge vector fields,
which we now call gluons, are singlets in flavor. The $\mathbf{8}$ was what
Han and Nambu called $SU(3)^{\prime \prime}$ (which we now call
$SU(3)_{color}$) and states that the interaction between the quarks is
mediated by an octet of gluons.

Other solutions to the statistics paradox, all of which failed,
were (i) an antisymmetric ground
state, favored by Dalitz, (ii) the idea that quarks are not
real, so that their statistics is irrelevant, and (iii) other atomic models.
Adding $q \bar{q}$ pairs
leads to unseen ``exploding $SU(3)$ states.''

The original version of the quark model did not consider ``saturation,''
why only the
combinations $qqq$ and $q \bar{q}$ occur in nature. In 1966 Daniel
Zwanziger and I surveyed the existing models and constructed new models
to see which models account for saturation~\cite{grezwa}. 
The only models that worked
were the parafermi model, order 3, and the equivalent 3 triplet or color
$SU(3)$ models. The states that are bosons or fermions in the parafermi
model, order 3, are in 1-to-1 correspondence with the states that are
color singlets in the $SU(3)$ model. Thus the parastatistics and explicit
color models are equivalent as a classification symmetry.

Some properties beyond classification agree in both models. The
$\pi \rightarrow \gamma \gamma$ decay rate and the ratio
$\sigma(e^+ e^- \rightarrow hadrons)/\sigma(e^+ e^- \rightarrow \mu^+ \mu^-)$
are the same in both cases, because it does not matter whether the
quark lines in intermediate states represent Green component quarks or
color quarks.

Properties that require gauge theory include (i) confinement, discussed
by Weinberg, by Gross and Wilczek and by
Harald Fritzsch, Gell-Mann, and Heinrich Leutwyler in 1973, which
explains why isolated quarks are not observed,
(ii) asymptotic freedom, found by David
Politzer and by Gross and Wilczek in 1973, which reconciles the 
quasi-free quarks of the parton
model with the confined quarks of low-energy hadrons, (iii) running of
coupling constants and precision tests of QCD, (iv) jets in high-energy
collisions, among other things.

\section{Summary}
The discovery of color resolved a paradox: quarks as spin-1/2 particles should
obey fermi statistics according to the spin-statistics theorem and should occur
in {\em antisymmetric} states; however they occur in the {\em symmetric} 
$\mathbf{56}$ of
the G\"ursey-Radicati $SU(6)$ theory. I resolved this paradox in 1964
by introducing a
new 3-valued hidden charge degree of freedom, color, via the parafermi
model of quarks in which color appears as a classification
symmetry and a global quantum number. I used this model to predict correctly the 
spectroscopy of excited states of baryons.
The other facet of the strong interaction, gauged $SU(3)_{color}$, was
introduced as a local gauge theory
by Nambu and by Han and Nambu in 1965.

\bibliographystyle{unsrt}

\end{document}